\documentclass[preliminary]{eptcs}
\providecommand{\event}{ALP4IoT 2017} 
\usepackage{breakurl}             
\usepackage{underscore}           
\usepackage{graphicx}
\usepackage{caption}
\usepackage{refstyle}
\usepackage{amsmath}
\usepackage{amssymb}
\usepackage{csquotes}
\usepackage[T1]{fontenc} 
\usepackage[utf8]{inputenc}
\usepackage{textgreek}

\title{IoT Architectural Framework: Connection and Integration Framework for IoT Systems}
\author{Onoriode Uviase \qquad\qquad Gerald Kotonya
\institute{School of Computer Science and Communication\\
Lancaster University\\
Lancaster, UK}
\email{\quad o.uviase@lancaster.ac.uk \quad\qquad g.kotonya@lancaster.ac.uk}
}
\def\titlerunning{IoT Architectural Framework}
\def\authorrunning{O. Uviase \& G. Kotonya }
\begin{document}
\maketitle

\begin{abstract}
The proliferation of the Internet of Things (IoT) has since seen a growing interest in architectural design and adaptive frameworks to promote the connection between heterogeneous IoT devices and IoT systems. The most widely favoured software architecture in IoT is the Service Oriented Architecture (SOA), which aims to provide a loosely coupled systems to leverage the use and reuse of IoT services at the middle-ware layer, to minimise system integration problems. However, despite the flexibility offered by SOA, the challenges of integrating, scaling and ensuring resilience in IoT systems persist. One of the key causes of poor integration in IoT systems is the lack of an intelligent, connection-aware framework to support interaction in IoT systems. This paper reviews existing architectural frameworks for integrating IoT devices and identifies the key areas that require further research improvements. The paper concludes by proposing a possible solution based on microservice. The proposed IoT integration framework benefits from an intelligent API layer that employs an external \emph{service assembler, service auditor, service monitor and service router} component to coordinate service publishing, subscription, decoupling and \emph{service combination} within the architecture.
\end{abstract}

\section{Introduction}

The term Internet of Things (IoT) refers to a heterogeneous network of physical and virtual objects embedded with electronics, software, sensors and connectivity to enable objects to achieve greater value and service by exchanging data with other connected objects via the internet \cite{cassimally2013}. \enquote{Thing} in terms of IoT, may be a person with a heart monitor implant, a farm animal with a biochip transponder, a field operation robot that assists in a search and rescue mission or any other natural or man-made object that can be assigned an IP address and provided with the ability to transfer data and to interoperate within the existing Internet infrastructure \cite{stankovic2014}.

An example of IoT-enabled environment is an integrated transport system that can be dynamically routed and reorganized in response to changing traffic needs and conditions \cite{zanellabui2014}. In healthcare, IoT has been used to follow-up on patient recovery and to assess that against a number of parameters unique to the patient by the use of IoT enabled devices \cite{chenxu2014}. The data gathered can also be used to compare patient responses to treatment in different environmental contexts on a global scale. Smart IoT devices can also be used to monitor and control energy use. In agriculture and food production, IoT can be used to manage production by monitoring and tracking variables that influence food production such as weather, politico-economic indicators, natural disasters, consumption, crop and animal diseases, etc. \cite{liuwang2014}. In assisted living, a ubiquity of IoT devices and services can help to address the need for independent living for the growing numbers of people living with physical disability, long-term conditions, social and age-related concerns \cite{tsirmpas2015}. 

\begin{figure}[]
\includegraphics[width=0.95 \textwidth]{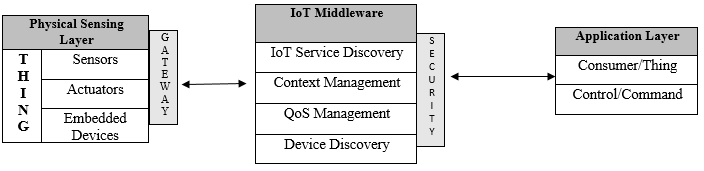}
\caption{IoT Basic Architecture}
\label{fig:basicarchitecture}
\end{figure}

\Figref{basicarchitecture} shows the basic IoT system architecture. The physical sensing layer contains embedded devices that make use of sensors to gather real world data. The gateway layer provides the mechanism and protocols for devices to expose their sensed data to the Internet (e.g. Wi-Fi, Ethernet, GSM, etc.). The middle-ware layer facilitates and manages the communication between the real world sensed activities and the application layer. The application layer maps onto applications that can be used by the consumer to send commands to real word objects over the Internet via mobile applications, webapps, etc.

IoT is a rapidly growing trend within major industries. In 2015, Samsung anticipated that 90\% of its products will be IoT enabled by 2017 \cite{samsunguk2015} and its entire product by 2020. Gartner also envisions an accelerated increase in IoT devices estimating about 21 billion of devices being IoT enabled by 2020 \cite{gartner}. The size and heterogeneity of the predicted growth adds extra complexity and urgency to the extant connection and integration problem \cite{buyyadastjerdi2016}.  Furthermore, IoT systems are likely to be distributed across different application domains and geographical locations creating hidden dependencies across domains, platforms and services. This has serious implications on how IoT systems are developed and evolved. Thus, the need for an intelligent, connection-aware framework has become a necessity. 

Service Oriented Architecture (SOA) offers a powerful framework for supporting the connectivity, interoperability and integration in IoT systems, it forms the backbone of present day IoT frameworks. While SOA goals are to primarily enhance IoT application interoperability, its monolithic usage in recent IoT frameworks further amplifies the problem of scalability especially with the enormous number of predicted \enquote{\textit{things}}. IoT systems tend to expand and with time, a capable SOA framework becomes too immovable to handle system extensibility. Microservice aims to fragment different IoT systems based on the System of Systems(SOS) paradigm to adequately cater for system evolution and extensibility. This paper reviews current IoT integration frameworks and proposes a partly explored, but promising solution to the problem of poor service integration, scalability, extensibility and fault tolerance in IoT.

This paper is organized as follows: Section 2 will contain the motivation and challenges in IoT architecture. In Section 3, we will present some requirements for IoT framework design and review a number of prominent IoT frameworks. Section 4 will discuss our approach to designing an IoT framework to support device connectivity, integration, interoperability, fault tolerance and scalability. In Section 5, we conclude our paper.

\section{Motivation and Challenges}

Interoperability refers to the ability of disparate IoT devices and services to exchange information and to use the information that has been exchanged. Current IoT initiatives focus largely on applications and devices that address disparate needs, but provide little scope for interoperation and connection. They lack a shared set of terms for describing services, service constraints, and service strategies. This reduces overall system visibility, creates redundant functionality, redundant services, presents problems for change management and limits the scope for resource coordination and reuse. IoT solutions need to support connection and interoperability as a \textit {building block} to provide mechanisms, processes and security that enable disparate devices and services to be joined up. The benefit is that IoT devices and services can interact with each other and create greater efficiencies, improve ease of use, provide better capability, greater choice, lead to economies of scale and potentially lower unit cost \cite{kiljander2014semantic}. Poor interoperability and connection are exacerbated by a number of factors including:

\begin{itemize}
\item \textit {Lack of standardized description of services}. Presently, there is still the lack of standards for naming convention of IoT device services, data information and description. This has posed a great challenge in the issue of integration. Recent trends in context awareness and data management have applied the semantic web technologies of Ontology Web Language (OWL) but still, there exist no standard methods to describe IoT services both at the high and low semantic level \cite{guillemin2015internet}. We have thus seen these naming conventions been disparate as a result of cultural, social backgrounds or coding habits whereby different terms are used to represent similar or the same entity. The solution of integrating heterogeneous IoT will largely benefit from integrated modelling techniques to provide appropriate knowledge acquisition and representation of the IoT domain.

\item \textit {Poor context-awareness for services}. The lack of adequate semantics for appropriate service context description is still evident in recent IoT system design. Current IoT systems suffer from insufficient context awareness of services due to inexpertly modeled semantics proliferating various unevenly distributed ontologies and incoherent semantics for services \cite{vermesan2011internet}. For a proper context-aware data processing in IoT domain, new approach should be utilized to modelling and designing rule engines for services.

\item \textit {Poor device service classification}. Several IoT systems utilize cataloguing of device services based on the device categorization. This type of practice attributes service to devices based on the unique identifier of both devices and services \cite{xiaoguo2014}. These services are made discoverable if an inquest is made to the service or device identifier. In IoT systems, this practice has long aided device discovery and service discovery but from an M2M viewpoint, service discovery is still unconvincing thus, if a device fails and is absent from the system it logs it as an unavailable service even if the same service with different unique identifier is provisioned for by another device. There has to be a new approach to service classification to accommodate such dynamic discovery of services.

\item \textit {Poor information visualisation and analysis}. Current initiatives provide little scope for tailored data collection and visualisation. Tailored data visualisation and analysis from an object’s activities and environment can provide invaluable insight into the well-being and the continued adequacy of the system \cite{ghidini2010snviz}.

\end{itemize}

\section{IoT Frameworks}

For an IoT framework to be reliable and dependable, some minimal set of measures should be satisfied to achieve integration and interoperability in IoT. These frameworks span across the IoT research communities ranging from academic research to organisational research which focus on integrating things in IoT. Since IoT paradigm itself is still in evolving state, we propose a set of minimal measures to be satisfied by IoT frameworks for integration. These are:

\begin{itemize}
\item \textit {Contract decoupling}: An IoT system contains heterogeneous devices with disparate communication protocols. An integration framework should be competent enough to efficiently handle contract decoupling. Contract decoupling is the ability of service consumers and service producers to independently evolve without terminating the contract between them \cite{richards2015microservices}. For example, a service might be in a JSON format and the service consumer needs an input in XML. The framework should provide support to transform the message to the format that fulfils the contract between them.

\item \textit {Scalability}: Given the evolving nature of IoT and the predictions and calculations by \cite{samsunguk2015} and \cite{gartner}, an efficient integration framework should be scalable and evolvable enough to support the billions of things soon to be connected to the internet. 

\item \textit {Ease of testing}: An integration framework should support ease of testing and debugging. It should provide support for debugging defects and failures, integration testing, component testing, system testing, compatibility testing, installation test, functional and non-functional testing, performance testing and security testing.

\item \textit {Ease of development}: An IoT integration framework should provide a means of easy development for developers. The framework should exclude all complexities and provide proper documentation for non-developers and developers with basic programming knowledge to easily understand the internals of the framework.

\item \textit {Fault tolerance}: An IoT system has to be dependable and resilient. An intelligent integration framework should effectively handle faults as IoT devices can eventually toggle between offline and online states. The framework should provide self-healing mechanisms for transient faults (network faults, node level faults, etc.), unauthorised access error, server crash failure, omission failure (when the server does not receive incoming requests from client), timing fault, etc.

\item \textit {Lightweight implementation}: Integration frameworks should have a lightweight overhead both in its development and deployment stage. It should be lightweight and easy to install, uninstall, activate, deactivate, update, versioning and adaptable.

\item \textit {Service coordination}: Service coordination is the orchestration and choreography of services. Service orchestration is the coordination of multiple services by a mediator acting as a centralised component. Service choreography on the other hand, is the chaining of services together to execute a particular transaction. Integration frameworks should support at least either or both to achieve reliability.

\item \textit {Inter domain operability}: The framework should further be extensible to support inter domain communication. For example, in a smart car domain, an integration framework should also provide support for communication and interaction with traffic lights, road closure, etc. belonging to a smart city domain.

\end{itemize}

\noindent
Regardless of the research community or disparity in research, they all aim to achieve extensibility, flexibility, scalability, design reuse and implementation reuse. The next sub-sections will present an overview of some IoT frameworks.

\subsection{Eclipse Smarthome Framework}

The Eclipse Smart Home (ESH) framework is designed for ease of resolution of IoT system and problems by developers who benefits from its interfaces, automation rules, persistence mechanism, and its SOA implementation \cite{eclipsesmarthome}. The ESH is a connection and integration framework for IoT smart home domain and it is independent of the connectivity features of the hardware but rather, makes emphasis on the implementation of a \textit{connector} to the framework. This connector is called \textit{binding} and is expected to be implemented at least once and only once for a particular communication protocol. The ESH has become largely famous as it is open sourced and thus, widely implemented as a smart home solution by a large market. This has contributed to its extensive number of shared API available for a range of commercial products \cite{eclipsesmarthome}.

\noindent
The ESH is explicitly fixated on home automation and it is built on five major stacks \cite{eclipseiot}. These are:

\begin{itemize}
\item \textit {Operating System}: This is the core support for the basic functions of the computer. The ESH equably runs on Linux (Ubuntu, Yocto-based Linux distribution), macOS and Windows.

\item \textit{The Application Container or Runtime Environment}: It is fully written in Java and uses the OSGi Runtimes (Eclipse Equinox) together with Apache Karaf and bundles it with Jetty HTTP server.

\item \textit{Communication and Connectivity}: The wide acceptance of ESH has seen its vast implementation thus, providing connectivity and communication between numerous off-the-shelf home automation products. Examples are, Belkin WeMo, Philips Hue, Sonos, etc. As its focus is specifically to home automation, it therefore provides support for offline communication capabilities within the ''Intranet of Things'' paradigm.

\item \textit{Data Management and Messaging}: The SOA approach of the ESH framework sees its implementation of an internal ''\textit{Event Bus}''. This bus is accessed and exposed externally through implemented protocol binding e.g. SSE or MQTT. The structure of the ESH also provides a persistence mechanism for database storage and also a rule engine to orchestrate runtime behaviour of things.

\item \textit{Remote Management}: The ESH framework is designed to make provision for remote monitoring, firmware update and configuration of connected devices.

\end{itemize}

\noindent
The ESH primarily functions as a piece of software incorporated into hardware to provide a collective point for coordination of the connectivity of things between each other and to an external network. The ESH framework was incorporated and implemented by openHAB to provide open source software to ease IoT application development.

\subsection{Calvin Framework}

The Calvin framework \cite{persson2015} is a hybrid framework of both the IoT and Cloud programming models to explicate the complexity of distributed computing, diverse programming languages and communication protocols. The framework is developed by consolidating theories from \textit{Actor model} and \textit{Flow Based Computing}. It divided the IoT application development into four separate and sequentially executed aspects which are:

\begin{itemize}
\item \textit{Describe}: In a primitive sense, Calvin characterises an \textit{actor} to consist of actions, communication ports and conditions that could trigger an action. These actors communicate with each other via the communication ports by exchange of tokens. In Calvin, the actors are software components mirroring: devices, services and computations. In the ''describe'' aspect, developers detail and express the actions, input/output relations and the preconditions required to trigger an action of an actor. It is also necessary in this aspect to prioritise the order between actions while separating the actor from the resources it represents in other to enhance the speed of shifting an actor from one runtime to another. 

\item \textit{Connect}: The next phase is to connect the ports of the already described actors by using a lightweight intuitive and declarative language called the \textit{CalvinScript}. The parameters needed by the actors are included and their pattern of connection initialised in this aspect

\item \textit{Deploy}: This aspect focus on the deployment of the described and connected component in a distributed runtime environment. Actors described and connected are lightweight thereby enabling migration between a mesh network of runtimes. The ''deploy'' aspect improvises a simple deployment pattern by passing the application script of an intended application for deployment into the runtime environment. The application script is always deployed to the closest accessible runtime. The runtime is then instantiated and all actors connected locally. Due to the distributed nature of the implemented runtime environment, actors can manoeuvre across accessible runtimes based on: locality, performance requirements, connectivity requirements and resources. In this aspect, the prescribed deployment algorithm takes into consideration some factors that may affect or influence the performance of the application, workload on existing actors, network congestion and runtime overload.

\item \textit{Manage}: This aspect involves the management of migrating actors between runtimes, error recovery, scaling, resource usage and updates by the distributed execution environment.

\end{itemize}

\noindent
The Calvin framework combines both IoT and Cloud models by designing the IoT system to leverage on the Cloud system to perform complex calculations and computations which are rarely possible with resource constrained actors. The framework thus prescribes a runtime API for connection and communication between runtimes and actors. The Calvin framework model propounds a distributed runtime environment and exhibits multi-tenancy as actors can share runtimes with actors from other applications. It also supports restrictions of high resource consuming actors at the \enquote{\textit{manage}} aspect. For example, image processing actors may sometimes be fully restricted or limited from a runtime.

\subsection{SOCRADES}

SOCRADES \cite{souza2008} is a service –oriented based integration architecture which provides generic components to aid the modelling of well detailed processes. It targets smart objects in manufacturing domain, which represents their behaviour as web services so as to enhance their capabilities. SOCRADES incorporates the pattern, concept and code of SIRENA \cite{bohn2006sirena} (a European Union funded project) to propose and design an integration infrastructure for web service and a framework for device supervision and life cycle which flawed in SIRENA.

SOCRADES is composed of four main parts which are the \textit{Device layer}, \textit{SOCRADES middle-ware} (comprising of two sub parts, \textit{Application part} and \textit{Device Service part}), \textit{xMII} and \textit{Enterprise application}.

\begin{itemize}
\item \textit{Device Layer}: This layer consists of devices that are web service enabled and they connect to the SOCRADES middle-ware layer using the Device Profile for Web Service (DPWS) model.

\item \textit{SOCRADES Middle-ware}: Devices in the device layer can connect to the Enterprise Application as they support web service. The SOCRADES middle-ware serves as a bridge between the device layer and the enterprise application. The main aim of this component is to simplify the management of devices in the device layer. Some other features and components of the SOCRADES middle-ware are: access to devices, service discovery, device supervision, service life cycle management, cross-layer service catalogue, and security support (optional). The SOCRADES middle-ware component also extends the functionality of the xMII component to other components and vice versa.

\item \textit{xMII Component}: This component is a System Applications and Products(SAP) in Data processing product: SAP xApp Manufacturing Integration and Intelligence (SAP xMII). It has features for: non-web service device enabled connectivity, visualisation service, graphical modelling, execution of business rules and connectivity to earlier SAP software via specific protocols. This component is integrated into the GUI of the Enterprise Application in a mash-up form by generating rich web content to suit the purpose.

\item \textit{The Enterprise Application}: This component consists of a GUI which exposes the system to the users. It receives data from the xMII component via specific protocols or by integrating xMII web content with the Enterprise Application GUI.

\end{itemize}

\noindent
SOCRADES is practically a web service integration architecture designed to support device connectivity and integration into an enterprise application such as in an ERP system. It is concisely based on the SOA paradigm.

\subsection{AllJoyn}

AllJoyn \cite{alliance2016alljoyn} is an open source framework targeting the connection and integration of things irrespective of their communication module, operating system and manufacturers. The framework provides a proximal network discovery between devices by abstracting out the details of the physical transport and providing a simple-to-use API for connecting things. Thus, the complexity of discovering nearby devices is handled by creating sessions (multiple sessions, point-to-point or group sessions) between devices for secure communication between them. The AllJoyn framework consists of some implemented common services and interfaces used by developers to integrate variety of devices, things or apps. The framework relies optionally on cloud services as it runs on a local network. This enables devices and apps to communicate within the network with only one gateway agent designed to connect to the internet. This in turn reduces security threat and the number of devices exposed to the internet threats. The framework comprises of two major components: \textit{AllJoyn Apps} and \textit{AllJoyn Routers}. These components can both dwell on the same or different physical device.

\begin{itemize}
\item \textit{AllJoyn Apps}: The Apps is a component of the framework that communicates directly with the AllJoyn Router and communicates with other Apps via the router. The AllJoyn App comprises of sub components which are the: \textit{AllJoyn App Code, AllJoyn Service Framework} and \textit{AllJoyn Core Library}.

\begin{itemize}
\item \textit{AllJoyn App Code}: This holds the logic for the AllJoyn Apps. It provides access to the AllJoyn Core API by connecting it with the AllJoyn Service framework or the AllJoyn Core Library components to provide such access.

\item \textit{AllJoyn Service Framework}: This is the component that implements the common services such as on-boarding a new device for the first time, sending notifications and controlling a device. These services allow the communication and interoperation between apps and devices.

\item \textit{AllJoyn Core Library}: This component provides access to the API for interaction with the AllJoyn network. It provides support for: session creation, object creation and handling, interface definition of methods, properties and signals and service/device discovery and advertisement.

\end{itemize}

\item \textit{AllJoyn Routers}: This component facilitates the communication between different Apps component. The AllJoyn framework comprises of three common pattern of communication between Apps and Routers.

\begin{itemize}
\item \textit{Bundled Router}: Here, the Router is bundled with an App based on a one to one relationship.

\item \textit{Standalone Router}: The Router is run as a standalone process on a system and it allows the connection to multiple Apps on the same device as the Router. This ensures that devices consume less overall resources.

\item \textit{Router on a different device}: The Router is run on a different device allowing connections from Apps. This is used for embedded devices that are resource constrained and thus, will utilise the ''\textit{Thin}'' version of the AllJoyn framework so as to contend with their limited resources.

\end{itemize}

\end{itemize}

\noindent
The AllJoyn framework concisely pursues a common ground for the connection, interaction and integration of things, devices and apps regardless of their individual OS, programming language and manufacturers.

\subsection{FRASAD}

The FRASAD \cite{nguyen2015frasad} is a development framework aimed at allowing developers design their IoT applications using sensor node domain concepts. This concept is model driven and as such the application code is generated from the designed model via a transformation process. The FRASAD (\textbf{FRA}mework for \textbf{S}ensor \textbf{A}pplication \textbf{D}evelopment) framework is an extension of \cite{thang2011model} by the addition and integration of two layers to the existing sensor node architecture. These two additional layers are the \textit{Application Layer} (APL) and the \textit{Operating System Abstraction Layer} (OAL). The essence of these two layers is to magnify the level of abstraction and thus concealing the lower levels. To achieve this, the framework employs the use of a robustly designed Domain Specific Language (DSL) to model the sensor nodes and separate the operating system from the application. The OAL is then contracted to explicate the modelled application based on the specific operating system for implementation. The OAL can be seen as an application generator: its core function is to generate the application code to be deployed in the targeted platform.
The FRASAD framework inherently follows the \textit{Model Driven Architecture} (MDA) approach by adopting three models/levels of abstraction. They are:

\begin{itemize}
\item \textit{The Computation Independent Model} (CIM): used to represent the actual information without exposing the structure of the system or the technology used for its implementation.

\item \textit{The Platform Independent Model} (PIM): This is the model that houses the application logics and requirements.

\item \textit{The Platform Specific Model} (PSM): The PIM is translated to the PSM based on the specific operating system of implementation, using a specifically designed DSL for mapping processes. The PSM is operating system specific and may use languages supported by the operating system.

\end{itemize}

\noindent
FRASAD in summary, is a framework which uses a multi-layered MDA with interaction between layers via some predefined interface. It utilises the node-centric model to facilitate the programming of individual sensor nodes using its rule-based programming model.

\subsection{ARIoT}

The ARIoT framework \cite{jo2016ariot}, consolidates the technology of \textit{Augmented Reality} (AR) by slightly extending the IoT infrastructure.  The core of the framework is to: dynamically identify IoT targeted objects within proximity of an IoT space, recognition and tracking of IoT objects feature information (examples are, object id, attribute values, object manufacturers, etc.) and augmenting the  interactive contents of the IoT object as services to a client. 

In this framework, IoT objects are set up to hold feature information and contents for services. An AR client system (e.g mobile phone) employs the AR technology to detect nearby IoT objects that meets the set-up criteria, through a defined standard wireless service discovery protocols. After the IoT object is identified via a standard protocol, there is message exchange between the AR client and IoT objects to further identify and filter out objects based on their distance and perspective (relative distance from other objects or their direction relative to the AR client). The AR client can then use the features and contents information from the exchanged message to track targeted objects, and augment the \textit{contents} on the clients' display screen. Individual IoT objects are independently attributed to their own recognition and tracking mode i.e. IoT object 1 may send information for Speeded Up Robust Features (SURF) algorithm for tracking and IoT object 2 can use a different model/algorithm for tracking. 

In summary, ARIoT integrates AR technology into the IoT infrastructure to provide a much friendlier environment to select target objects and subject them to control, interactions and tracking. The AR client is mostly a mobile device which contains an implemented algorithm to identify, recognise and track IoT objects via predefined standard protocols. This framework improves on scalability by eliminating the use of a central server via AR client, allowing IoT objects to directly communicate and exchange information on a need to know basis.

\subsection{AVIoT}

The AVIoT \cite{jeong2015aviot} is an IoT framework for visualising and managing IoT objects present in a specific environment (e.g. a smart home). This framework aims to allow users to apply its web-based visual authoring tools to abstract and program the behaviour of IoT things. Hence, end users can easily monitor and define the behaviour of IoT things without prior internal knowledge of the architecture or connection system of the sensors. The AVIoT framework is proposed to allow visual configuration and management of things within an IoT environment with ease. This framework follows a principled process of: Abstraction of Sensors and Actuators, Interactive IoT Visualisation, and Interactive IoT Authoring.

\begin{itemize}
\item \textit{Abstraction of Sensors and Actuators}: In this process, physical things that are present in an IoT environment are abstracted and virtualised to interact with other virtual things by defining it as a node. This node will be defined based on the generic features of things such as name, type of thing, type of visualisation, position, children (if any) and functional string behaviour (contains a script code which can be assessed at runtime). Things are abstracted and virtualised to contain \textit{physical sensors, virtual sensors, virtual actuators} and \textit{physical actuators}. While physical sensors and actuators are the real-world things, the virtual sensors core function is to detect events of significance which the end user has expressed interest in. The virtual actuators' core function is to adapt the end user pre-defined behaviour to the physical actuators. The virtual actuators do not hold values like the virtual sensors, rather, they trigger the actions of the physical actuators. The interaction between the abstracted components follows the order of \textit{physical sensor -\textgreater virtual sensor -\textgreater virtual actuator -\textgreater physical actuator}.

\item \textit{Interactive IoT Visualisation}: This is the visualisation of virtual things within the IoT environment to promote the management of physical things. This phase uses a \textit{client} – web browser or web application and a \textit{server} to facilitate the visualisation and management of virtual things. The assumption is that the server has identified the devices and holds information on their various connectivity modules. The server is also used as storage for information with regards to visualisation and authoring of things. This phase exploit REST as the interface for communication between the client and server. For 3D visualisation, the server holds a virtual plan of an indoor environment in their 3D model state. This can be queried by a running client to provide a 3D visualisation of things at runtime. The 3D visuals are rendered like wire frames and visual icons are displayed based on the location of the physical things to identify sensors and actuators.

\item \textit{Interactive IoT Authoring}: In this phase, virtual things can be edited, added, or deleted from the given set of physical things. Things can also be relocated by the end user to a different position and rediscovered in the new location by the server. The physical sensor interacts with the real world and outputs observed numeric data which is received by the virtual sensor. The major focus of the virtual sensors is on events (pre-configured) that are of concern to the end user. This information is translated and passed to the virtual actuator which then acts by triggering the appropriate action of the physical actuators.

\end{itemize}

\noindent
The AVIoT is an IoT framework for visualisation and authoring of things in a smart environment by the end users. It is implemented on the web and fully focused on indoor IoT environments. The framework allows for end users to abstract and define things in order of hierarchy by utilising its web application tools for visualisation and authoring.

\subsection{Summary}

Each of the IoT frameworks discussed, have applied various software architectures (MDA, SOA, etc.). The most prominent and productive architecture can be seen to be the SOA. SOA is somewhat beneficial in IoT integration frameworks as it provides contract decoupling and heterogeneous protocol sceptic interoperability.The frameworks reviewed have their disadvantages and advantages which are either attributed to the selected architecture to provide an IoT solution or the pattern of implementation. For example, The ESH has its strengths in providing support for heterogeneous communication protocols while providing means of creating \enquote{\textit{bindings}} for new protocols. It supports a wide range of off-the-shelf products, it is easy to develop and deploy. Regardless, the ESH is limited to smart home, it depends on an enterprise bus for message transport which may become a single point of failure, It has no support for merging services for example, if a device X requires a composite service of temperature and humidity, it has to make individual request for each of these services especially if the services is available on two different devices Y and Z. The ESH components are monolithically designed and makes it hard enough to fit into a larger system as a sub-system. 

The obvious strength of the Calvin framework are, its flexibility due to the use of a cloud system. Despite this strength, the Calvin framework shows some limitations as, developers need to learn a new sketchy language, there is no service combination, no provision for service management and there is high threat to system security e.g. as actors migrate between runtimes, there is no sonsideration for security threat factor. Hence an actor whose security has been compromised will pose a threat to several other actors on different runtimes.

The SOCRADES framework provides support for heterogeneous system, its Enterprise Application component presents users with an easy to use GUI thus reducing deployment time and providing ease of system management. Nonetheless, single point failure is present in its use of an enterprise service bus, there is no intrinsic knowledge of the services and it lacks support for merging fine-grained services to coarsed services on demand.

Decisively, the monolithic implementation of SOA, its evident single point of failure and complexity in architecture due to the billions of things to provision for, further complicates the question of scalability, fault tolerance and reliability. An IoT connection and integration framework needs to be scalable enough to support billions of things connected, it should tolerate common faults such as latency and provide a means for identifying, merging, recreating and granting secured remote access to these services. 

\section{Proposed Approach}

In building an accomplished scalable, adaptive and fault tolerant framework for IoT, the distributed systems architectural style is a favourable approach. This way, software components are remotely accessed, via various remote access protocols. This architectural style consists of several patterns which include: \textit{Client-server (2-tier, 3-tier or n-tier), Peer-to-Peer, Object request broker, Space based architecture, Shared nothing architecture, REST, Service-Oriented, Cloud computing pattern,} etc. To attain a scalable, evolvable and fault tolerant IoT framework, the software architectural patterns currently utilised, should be improved to accommodate the IoT paradigm and fufil the presented minimal requirements. We propose the adoption of \textit{Microservices} which is a service based (service-oriented pattern) architecture. Microservices are characterised by the use of simple API which are thin layered (light weighted compared to SOA). While sceptics may argue microservices are same as SOA, the similarity is that they are both service based architecture with emphasis on service use and reuse. They differ in terms of their \textit{architectural style, architectural characteristics, service characteristics} and \textit{capabilities}. The term service in the context of SOA, is defined by \cite{mackenzie2006reference}, as ''\textit{a mechanism to enable access to one or more capabilities, where the access is provided using a prescribed interface and is exercised consistent with constraints and policies as specified by the service description}''. This definition defines services based on its ownership, accessibility and qualitative attributes. However, services could be further defined based on their \textit{atomity} (size of services) and interaction (in terms of service unification). Service atomity and classification, presents a great distinctive feature between SOA and microservices. Interactions between different services are conveniently handled in microservices architecture and as such will be advantageous in an intrinsically asynchronous IoT system.

\subsection{Microservices for IoT}

The main aim of microservices is to coordinate distributed applications as a collective implementation of services running autonomously and independently based on its described process context \cite{richards2015microservices}. The growth in technology has seen IoT devices evolve with different communication protocols, making it difficult to achieve connectivity, scalability and integration especially in monolithically fabricated frameworks. Another challenge is the transparency of the support for legacy protocols with or without an upgrade. With the need to reduce power consumption to achieve an eco-friendly envinroment, IoT devices has been designed to consume less energy and as such, limiting its resources (storage, processing capacity, etc.). This has also become a challenge in IoT from their device perspective. Devices are expected to run for a prolonged period, producing and consuming services, exchanging data and high payload, etc. All these core fuctionalities will in turn impair their battery life.

Microservices architecture (share-as-little-as-possible) presents the possibility to resolve these challenges as it promotes independent service deployment, service atomity , deterrent to single point failure, augment security transaction management and service choreography. The next sub-sections present an envisioned design of an IoT system in microservices architecture, based on some complex challenging issues in IoT.

\subsubsection{Service Identification and Classification}

Microservices classifies services into 

\begin{itemize}
\item \textit{Functional Services}: services that support the operational functions of a smart system in an IoT domain. In IoT, these services are mostly literals (i.e. numbers, letters, etc.). These services are exposed to be utilised by an external system or device.

\item \textit{Non-Functional Services}: services that are related to non-operational tasks (such as: Logging, authentication, monitoring, authoring, auditing, etc.) but must be utilised for a reliable operation of the system. These services are not exposed and are utilised by a system or device wishing to integrate into an existing IoT system.

\end{itemize}

\noindent
The concept of service taxonomy (service description) as viewed in microservices architecture, is adopted into IoT domain to reproduce an IoT architectural framework to facilitate service identification and classification.

\begin{figure}[h]
\centering
\includegraphics{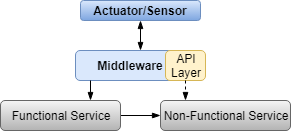}
\caption{Microservice for IoT Service Identification and Classification}
\label{fig:identificationservice}
\end{figure}

\noindent
Functional and non-functional services are clearly identified and separated. The API layer receives request and can trigger the non-functional services on its own or via the functional services to hide the non-functional service from the real world.

\subsubsection{Service Atomity}

The fundamental concept of microservices is its fine-grained services which are single-purposed and thus perform optimally in executing that one functionality. In IoT, fine-grained services are common especially among devices with communication protocol like the MQTT, AMQP etc. One challenge in IoT is latency issue which sometimes emerge as a consequence of large, coarse-grained services been consumed by IoT devices of limited resources. In microservices, the impact of fine-grained services aids in software development, testing, deployment and maintenance. The IoT system can immensely benefit from adopting this pattern of atomized services. Building minute services accurately in microservice architecture is somewhat complex and similar problem can be faced when adapting it into IoT.

\begin{figure}[h]
\centering
\includegraphics[width=0.75\textwidth]{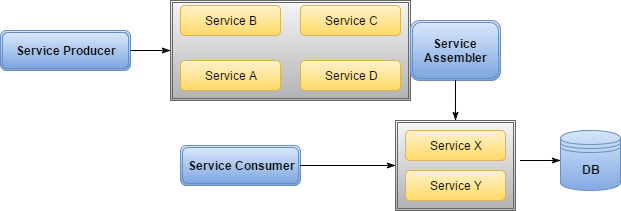}
\caption{Microservice for IoT Service Atomity}
\label{fig:serviceatomity}
\end{figure}

\noindent
\Figref{serviceatomity} is bolstered to fit into IoT and expel the problem associated with granular services. In microservices, the service granularity problem hinders performance and transaction management. If a transaction requires services A, B and C, it takes the sum of $T_A$, $T_B$ and $T_C$ (where $T_N$ is the time to process service N) time to process. Thus, if it takes 200ms to process each service, then it will take 600ms to process the transaction. A \emph{Service Assembler} is introduced to handle consumer request and service coupling (combining two or more services A, B and/or C to generate a new functional service X which adheres to the service contract of the consumer based on the implementation details). With the introduction of the service assembler component, consumer request for composite service transaction is processed and dispensed by the assembler, thereby reducing the time spent on remote access call. The service assembler basically combines fine grained services to a composite service which can be advertised as a separate service depending on the amount of request for similar services or transaction.

\subsubsection{Service Coordination}

Service coordination can be looked at as a systematic way by which services are arranged, allocated and analysed. It is based on two main service communication concepts which are \textit{Service Choreography} and \textit{Service Orchestration}.  Service orchestration is the organisation of various services collectively by a centralised component which can be seen as the \emph{mediator}. The \textit{mediator} coordinates different services that are necessary to complete a particular transaction. Service orchestration is mostly dominant in SOA, hence, IoT frameworks adopting the SOA paradigm utilise an integration hub or a middle-ware in the overall architecture to provide this functionality. The deficiency in this approach is that the system may experience lags in multiple service requests required to complete a single transaction. In SOA, systems mostly rely on the combination of multiple composite (coarse-grained) services to complete a single business request. In an IoT system, services are commonly fine-grained and as such, service choreography will be enough for small scale IoT systems. 

\begin{figure}[h]
\centering
\includegraphics[width=0.75\textwidth]{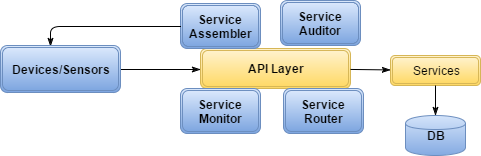}
\caption{IoT Consolidated Service Topology}
\label{fig:servicecoordination}
\end{figure}

As commonly used in microservices, service choreography is the organisation of several services to execute a single transaction without a central mediator. This initiates a chain reaction or service chaining, where one service calls another service and the other service calls another until the last required service is called. While this is inexpedient as it will increase the remote access call time for multiple service requests, the addition of an autonomous service assembler component will be beneficial in reducing the call time. Service chaining is uncomplicated in IoT considering its fine grained services which comprises mostly of literals. The service assembler functions in a manner to eliminate the need for a new remote service contract after \textit{service combination}. Other components like the service auditor, service monitor etc. are involved in the runtime operation of the service assembler as seen in \Figref{servicecoordination}

\subsubsection{Interoperability in IoT}

The IoT paradigm is built around service sharing and to achieve interoperability, IoT frameworks should allow service coupling and recoupling. In microservices, the API Layer is used rather than the Enterprise Service Bus (ESB) conventionally used in SOA. The benefit of this API approach to IoT is that it allows for the level of granularity of services to be changed without affecting the consumers. For instance, a coarse grained service can be dismantled into finer grains or vice versa to increase performance, scalability and ease of deployment of the system. With the API component, the consumers do not need to change service contract or adapt themselves. Rather, the API knows to call the split service when a call is made to the coarse service. IoT will do better to benefit from an implemented intelligent API layer. Unlike SOA middle-ware that utilises a mediator for service orchestration, message enhancement and protocol transformation implemented within the ESB, an intelligent API layer for IoT will make use of smaller components. These components are not enclosed within the API layer, but will be available at runtime to execute similar functionality e.g. service combination, \textit{contract decoupling} etc. These components will compensate for heterogeneous protocol support which lacks in a standard microservice architecture.

\begin{figure}[h]
\centering
\includegraphics[width=0.75\textwidth]{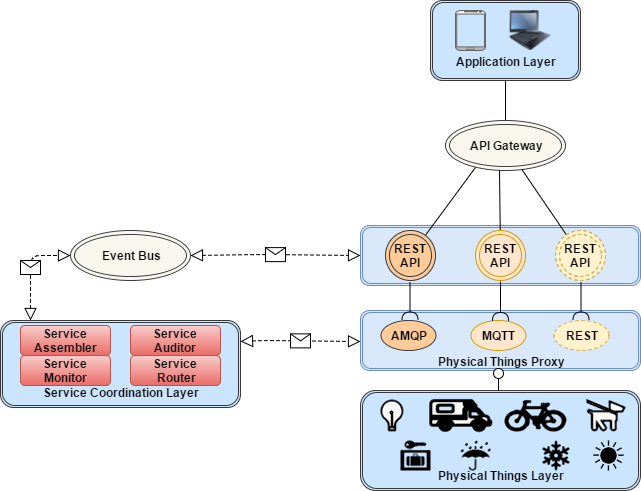}
\caption{Microservice for IoT Architecture}
\label{fig:microiotarch}
\end{figure}

\Figref{microiotarch} represents a Microservice architecture for IoT. The introduction of the \textit{service coordination layer} is to facilitate service identification, classification, combination and choreography. The figure shows several microservices implementing the various required interface (e.g MQTT, REST, etc.) which are provided by the physical sensing things. For example, if weather report is requested from the application layer, it makes a call to the API Gateway which then calls the appropriate REST API and the \textit{command} sent via the Event Bus. On receiving the command, the event bus transports the request to the service coordination layer (were necessary service identification, discovery, binding and routing is done). The command is channelled through the relevant protocol and the physical device acts accordingly. In the case of physically sensed activity, the event update is sent via the provided protocol to the service coordination layer. This event update is pushed via the event bus to a well-suited REST API and the update reflected on the mobile device.

\section{Conclusion and Future Work}

This paper has explored the lingering problems in integrating IoT devices and systems. It has reviewed some existing framework with much focus on the adapted SOA paradigm. This paper has further proposed an architectural framework for IoT which will also be advantageous from the developers’ point of view. As more devices gets connected to the internet, there is a need for a highly scalable, extensible and fault tolerant integration framework. Our belief is that adopting the microservice architecture in IoT to remodel integration frameworks, will enhance the reliability of IoT systems. The proposed framework is yet to be fully implemented and tested but it presents a direction of investigation for the IoT community on ways the problem of scalability, fault tolerance, ease of deployment, inter-domain communication, ease of development and lightweight implementation can be resolved. 

\bibliographystyle{eptcs}
\bibliography{mybiblography}
\end{document}